\documentclass[11pt]{IEEEtran}
\usepackage{amssymb}
\usepackage{hyperref}
\usepackage{amsmath}
\usepackage{graphicx}
\usepackage{soul}
\usepackage[font=small,labelfont=bf]{caption}

\input{Qcircuit}

\newsavebox{\mybox}

\newcommand{\eps}{\varepsilon}

\begin{document}

\title{Simulating Quantum Circuits by Shuffling Paulis}
\author{Patrick Rall, \textit{Quantum Information Center, University of Texas at Austin}\\ \today\\ \vspace{-10mm}}
\maketitle

\begin{abstract}
    Verification of NISQ era quantum devices demands fast classical simulation of large noisy quantum circuits. We present an algorithm based on the stabilizer formalism that can efficiently simulate noisy stabilizer circuits. Additionally, the protocol can efficiently simulate a large set of multi-qubit mixed states that are not mixtures of stabilizer states. The existence of these `bound states' was previously only known for odd-dimensional systems like qutrits. The algorithm also has the favorable property that circuits with depolarizing noise are simulated much faster than unitary circuits. This work builds upon a similar algorithm by Bennink et al. (Phys. Rev. A 95, 062337) and utilizes a framework by Pashayan et al. (Phys. Rev. Lett. 115, 070501). 
\end{abstract}

Quantum devices demand classical simulation for analysis and verification. Fast classical evaluation of Clifford circuits is a celebrated early accomplishment of quantum information theory \cite{tableau}. Using stabilizer techniques to simulate non-stabilizer quantum processes has attracted recent attention, since these algorithms can sometimes run in polynomial-time in the number of qubits and the circuit depth. 

Gate-based superconducting qubit devices are a target for classical analysis. Quantum error correction protocols exhibit deviation from Clifford circuits primarily due to noise, and are hence a candidate for near-stabilizer simulations. It remains to be seen if stabilizer strategies are viable for the analysis of other superconducting qubit protocols, since depolarization increases the ``Cliffordness'' of a channel. Pushing the limits of stabilizer simulation strategies is necessary since brute-force strategies become impractical at about 50 qubits. Even for smaller systems, benchmarking strategies for superconducting qubits rely on Clifford circuits \cite{benchmarking}. Techniques for noise analysis are shaped by what circuits are classically tractable. The runtime of classical simulations can also be used as a benchmark for quantum speedup.

If the gate set is fixed to Clifford+$T$ and no noise is present, \cite{gosset} achieves simulation with exceptional accuracy and efficiency. But noisy circuits demand support for mixed states, quantum channels and POVMs, as well as gate-set agnosticism to support different types of hardware. Quasiprobability representations admit such flexibility, permitting descriptions of any state or channel as well as providing precise measures of ``non-Cliffordness'' \cite{rbdobv16} \cite{resource}. These measures have been connected to contextuality \cite{rbdobv16} \cite{dovbr16}, so these measures in a sense quantify ``quantumness'' \cite{hwve14} \cite{rbdobv15}. 
In \emph{Section I} we present a randomized algorithm that can estimate outcomes of arbitrary noisy quantum circuits. We construct an upper bound for the runtime which proves that the protocol is efficient for probability distributions over Clifford circuits. In \emph{Section II} we show that there exist states that are \emph{not probability distributions over stabilizer states} that can be simulated efficiently. These were previously known to exist in odd-dimensional systems but not even dimensions. In \emph{Section III} we examine a particular quantum circuit to clarify some aspects of the protocol.

\newcommand{\benn}{Bennink et al. }
\newcommand{\pash}{Pashayan et al. }

\newcommand{\Tr}{\text{Tr}}
\newcommand{\Pau}{\mathcal{P}^n_+}
\newcommand{\Prob}{\mathsf{P}}

\newcommand{\abs}[1]{\left|#1\right|}
\newcommand{\expe}{\mathbb{E}}
\newcommand{\Var}{\text{Var}}
\newcommand{\brac}[1]{\left[#1\right]}
\newcommand{\pare}[1]{\left(#1\right)}
\newcommand{\nega}{\mathcal{D}}
\newcommand{\spac}{\hspace{5mm}}

\section{Algorithm Description}

Quasiprobability strategies for classical simulation of quantum circuits represent density matrices as weighted sums of a fixed set of hermitian operators. Some examples include projectors onto stabilizer states \cite{bennink} and the phase-point operators of the qutrit discrete Wigner function \cite{pash}\footnote{Actually, \cite{pash} describes a much more general framework, and then plugs in the qutrit Wigner function as an example.}. Our algorithm uses Pauli matrices, so the `quasiprobabilities' are really just Bloch vector components. 

Let $\{\sigma_i\}$ be $n$-qubit tensor products of Pauli matrices. Their normalized versions $\{\bar \sigma_i\}$ are trace-orthogonal and form a self-dual frame:
\begin{equation}
\begin{split}
    \bar \sigma_i = \frac{\sigma_i}{\sqrt{2^n}}; \spac & \Tr(\bar \sigma_i  \bar \sigma_j) = \delta_{ij} \\
    \rho = \sum_i \bar \sigma_i \Tr(\bar \sigma_i \rho) =& \sum_i \sigma_i \Tr\pare{\bar \sigma_i \frac{\rho}{\sqrt{2^n}}}
\end{split}
\end{equation}

Consider applying a quantum channel $\Lambda$ to a state $\rho$.
\begin{equation}
\Lambda(\rho) = \sum_i \Lambda( \sigma_i) \Tr\pare{ \bar \sigma_i\frac{\rho}{\sqrt{2^n}} }   
\end{equation}

The resulting state can be regarded as a weighted sum over $\Lambda( \sigma_i)$. The weights $\Tr\pare{\bar \sigma_i \rho/\sqrt{2^n}} = r_i$ define a scaled Bloch vector for $\rho$.

Consider iterating this process through several channels $\Lambda_1, ..., \Lambda_j, ..., \Lambda_m$. At every step, an operator $\rho_j = \Lambda_j( \sigma_i)$ is encountered.  $\rho_j$ can be decomposed into a normalized Bloch vector. Repeating this for every channel yields a sum with exponentially many terms, each corresponding to choice of $\sigma_i$ for each  $\rho_j$. Let $f$ be a function that encodes such a choice. Let $r_f$ be the product of the normalized Bloch vector components along the way, and $\sigma_f$ be the final Pauli matrix.
\begin{equation}
\Lambda_m(...\Lambda_1(\rho)...) = \sum_f r_f \sigma_f
\end{equation}

At the end of a quantum circuit an observable $E$ is measured. This observable could be a projector onto a state, a tensor product of Pauli matrices, or some other hermitian operator. The goal of the protocol is to estimate the expectation of this observable which we call $\Prob$:
\begin{equation}
\Prob = \Tr(E \Lambda_m(... \Lambda_1(\rho)  ... )  ) = \sum_f  r_f \Tr(E \sigma_f) 
\end{equation}

We construct an estimator for $\Prob$ called $\hat \Prob$, corresponding to choosing a random term in the sum:
\begin{equation}
\hat \Prob = \frac{r_f}{p_f} \Tr(E \sigma_f) \text{ with probability } p_f
\end{equation}

$\hat \Prob$ is an unbiased estimator for $\Prob$:
$$\expe\pare{\hat \Prob} = \sum_f p_f \frac{r_f}{p_f} \Tr(E \sigma_f) = \sum_f \Tr(E \sigma_f) r_f = \Prob $$

The runtime of the protocol depends on the rate of concentration of the average of $N$ samples from $\hat \Prob$. To upper-bound this we use the Hoeffding inequality \cite{hoeffding}:

\begin{equation}
\begin{split}
\Pr\pare{ \abs{ \sum_i  \frac{\hat \Prob_i}{N} - \Prob } \geq \eps } \leq 2 \exp\pare{ \frac{2N \eps^2}{ (\text{max } \hat \Prob - \text{min } \hat \Prob )^2} }  \\
 = \delta \text{ if } N \in O\pare{ (\text{max } \hat \Prob - \text{min } \hat \Prob )^2 \cdot \frac{1}{\eps^2} \log \frac{1}{\delta}   }
\end{split}
\end{equation}

To obtain $\text{max } \hat \Prob - \text{min } \hat \Prob$ we must pick a probability distribution $p_f$. Each choice of $\sigma_i$ from operators $\rho_j$ is made independently with probability:
\begin{equation}
    p_i = \frac{\abs{r_i}}{\nega(\rho_j)} \text{ where } \nega(\rho_j) = \sum_{k} \abs{r_k}
\end{equation}

If the circuit's input state $\rho$ is a tensor product $\rho_1 \otimes ... \otimes \rho_n$ then a Pauli matrix can be sampled from each $\rho_i$ individually. The procedure fully supports states and channels acting on only part of the circuit Hilbert space. 

\begin{equation} \frac{p_f}{r_f} = \frac{\prod_j  \abs{r_{j}} \big/\nega(\rho) }{ \prod_\rho  r_{j}} = \text{sgn}\pare{  \prod_j  r_{j} } \prod_j  \nega(\rho_j)\end{equation}
    \begin{equation} \text{max } \hat \Prob - \text{min } \hat \Prob \leq 2\cdot \max_f \abs{\Tr(\sigma_f E)}\cdot \prod_j  \nega(\rho_j) \end{equation}
\begin{equation}N \in O\pare{  \max_f \abs{\Tr(\sigma_f E) }^2 \cdot \prod_j  \nega(\rho_j)^2 \cdot \frac{1}{\eps^2} \log \frac{1}{\delta}} \label{eq:runtime} \end{equation}

The quantity $\nega(\rho) = \sum_{i} \abs{r_i}$ is reminiscent of the \emph{sum-negativity} of a quasiprobability distribution, a common measure of ``non-Cliffordness'' \cite{resource, hwve14, rbdobv15, rbdobv16, rbdobv16, dovbr16}. In our case, Bloch vectors are not really quasiprobabilities (since they do not always sum to 1), but some similarities remain. We will discuss $\nega$ in section II. For this section all that matters is that $\nega(\rho) \leq 1$ for stabilizer states, and that $\nega(\sigma_i) = 1$. 

Observe that if all operators $\rho_j$ encountered along the way satisfy $\nega(\rho_j) \leq 1$, then the $\prod_j  \nega(\rho_j)^2 $ term in (\ref{eq:runtime}) does not blow up. To ensure fast runtime in the number of qubits, $\max_f \abs{\Tr(\sigma_f E)}$ must also scale well. If $E = \ket{\psi}\bra{\psi}$ for some state $\ket{\psi}$, then $\max_f \abs{\Tr(\sigma_f E)} \leq 1 $ since the eigenvalues of $\sigma_f$ are $\pm 1$.

If however $E$ is a Pauli operator $\sigma_i$, then $\max_f \abs{\Tr(\sigma_f E)} = 2^n $ and the runtime is exponential in the number of qubits. If $E$ is a local observable like $\sigma_I^{\otimes n_1} \otimes E_\text{local} \otimes \sigma_I^{\otimes n_2} $, then the discarded qubits also cause a blowup of $2^{n_1 + n_2}$. If $E$ is of such a form then a different version of the algorithm must be used.

This is a problem unique to our protocol and is not an issue in \cite{bennink, pash}. Pauli matrices satisfy $\max_i \Tr(\sigma_i) = 2^n$, while stabilizer states $\ket{\phi_i}\bra{\phi_i}$ and Wigner function phase point operators $A_i$ satisfy $\Tr(\ket{\phi_i}\bra{\phi_i}) = \Tr(A_i) =1$ \cite{gross}. With Pauli matrices all of the trace is concentrated on the identity, so tracing out qubits causes blowup.

The procedure we described above forward-propagates the input state $\rho$ through the channels $\Lambda_1, ..., \Lambda_m$ and finally takes the expectation with $E$. One can also back-propagate $E$ through $\Lambda_m^{-1}, ..., \Lambda_1^{-1}$ and finally take the expectation with $\rho$. Since $\abs{\Tr(\sigma_f \rho)} \leq 1$ the blowup is removed.

To do this we must define $\Lambda^{-1}$. If $\Lambda$ is non-unitary then there does not exist a channel $\Lambda^{-1}$ such that $\Lambda^{-1}(\Lambda(\rho)) = \rho$. However, all we need is to define $\Lambda^{-1}(E)$ such that $\Tr( \Lambda(\rho) E ) = \Tr(\rho  \Lambda^{-1}(E))$. This is always possible:
\begin{equation}
\Lambda^{-1}(E) = \sum_j\bar \sigma_j  \Tr( \Lambda(\bar \sigma_j) E )
\end{equation}
Then:
\begin{equation}
\begin{split}
\Tr(\rho  \Lambda^{-1}(E)) = \Tr\bigg( \rho \sum_j \bar \sigma_j \Tr( \Lambda(\bar \sigma_j) E )   \bigg) \\
 = \Tr\bigg( \Lambda\bigg(\sum_j  \bar \sigma_j\Tr\pare{ \rho \bar \sigma_j }\bigg) E \bigg)=\Tr( \Lambda(\rho) E )  
\end{split}
\end{equation}

\begin{figure}[t]
    \begin{center}
        \includegraphics[width=0.3\textwidth]{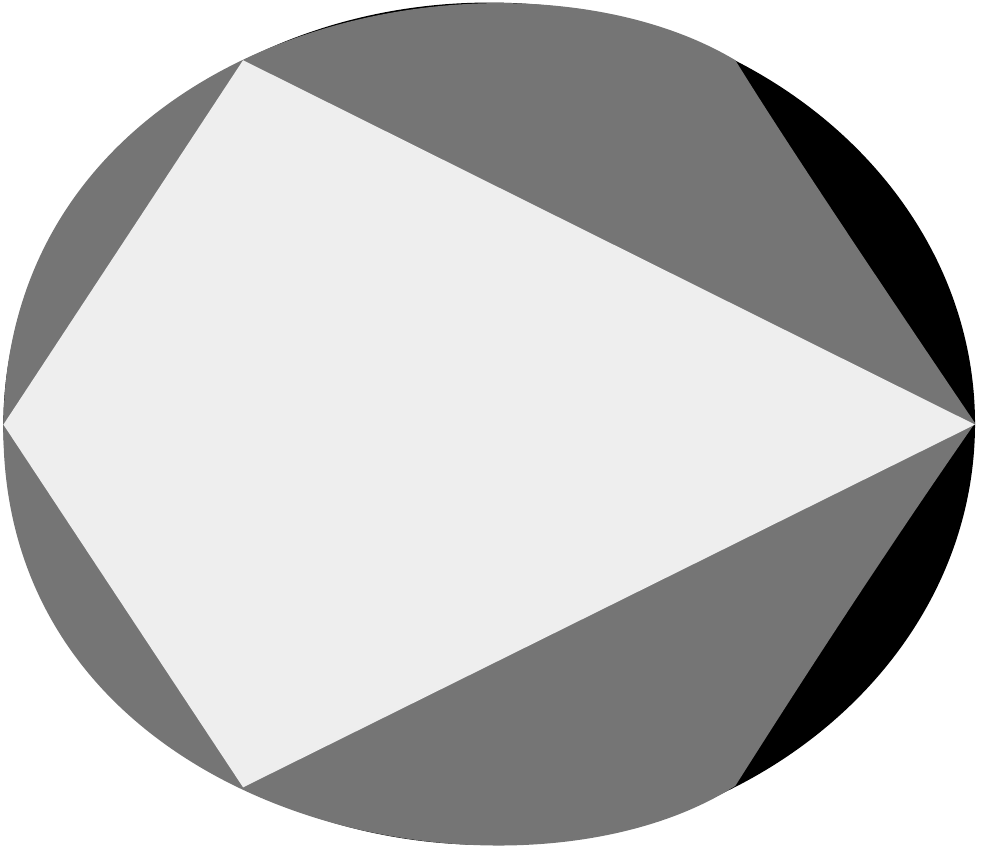}
        $$\text{a) } \rho = \frac{\sigma_{II}}{4} +x \pare{\sigma_{XX} + \sigma_{ZZ} - \sigma_{YY}} + y\pare{\sigma_{ZI} + \sigma_{IZ}}$$
        \vspace{3mm}

        \includegraphics[width=0.3\textwidth]{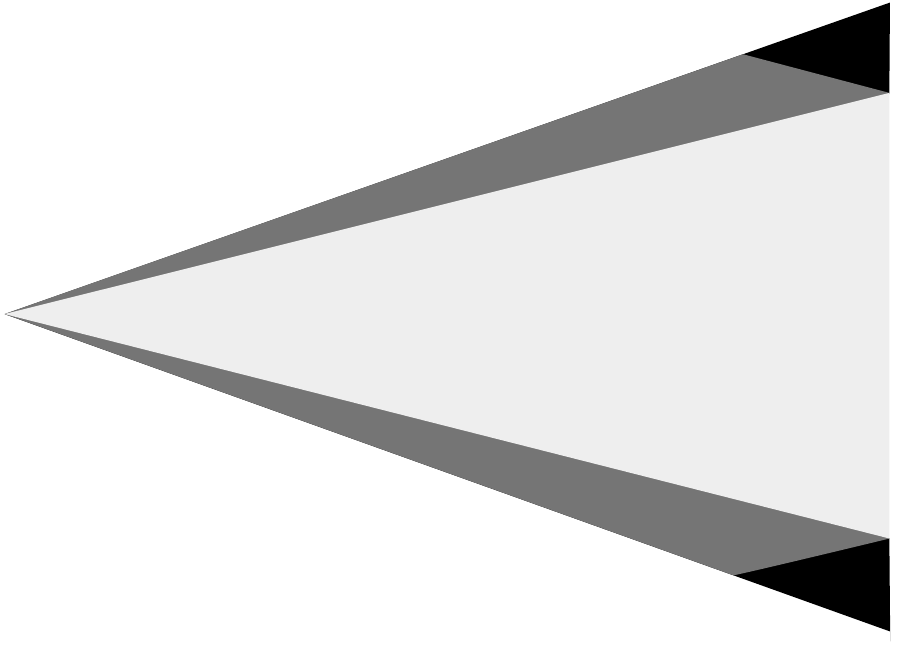}
        $$\text{b) } \rho = \frac{\sigma_{II}}{4} +x  \sigma_{ZZ}  + y\pare{\sigma_{XX} + \sigma_{XY} + \sigma_{YX} - \sigma_{YY}}$$
        \vspace{3mm}

        \includegraphics[width=0.3\textwidth]{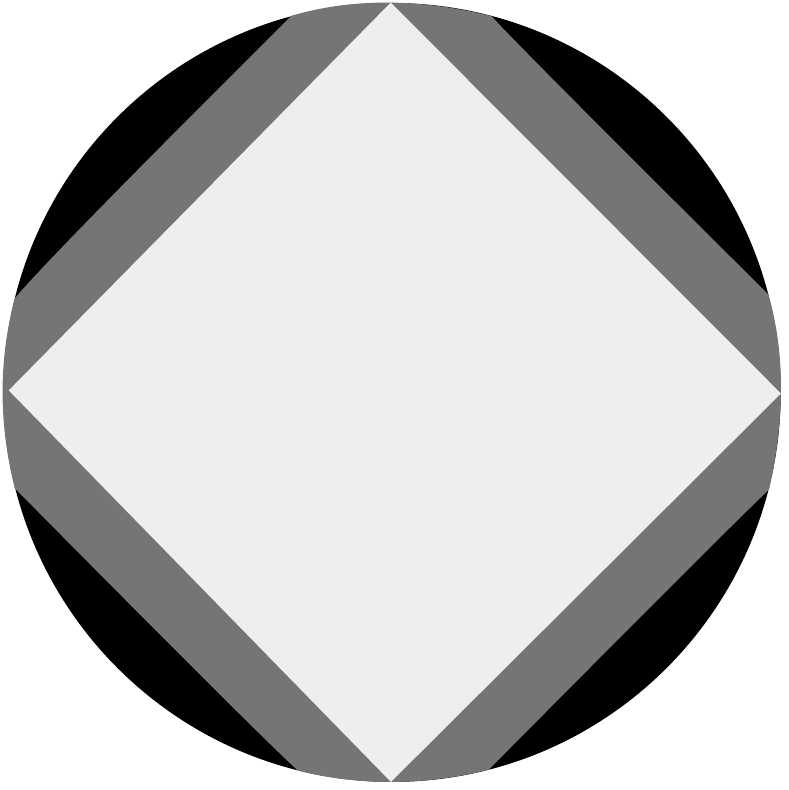}
        \vspace{2mm}
        $$\text{c) } \rho = \frac{\sigma_{II} + 0.8\sigma_{ZZ}}{4} +x \pare{\sigma_{XX} - \sigma_{YY}} + y\pare{\sigma_{XY} + \sigma_{YX}}$$
    \end{center}
    \caption{Some cross sections of two-qubit mixed states. Stabilizer mixtures are light gray, bound states are dark gray, and magic states are black. \label{fig:cross-section}}
    \vspace{-5mm}
\end{figure}

Since the input to the circuit is always a state, back-propagation always produces a fast runtime provided the observable $E$ and all operators $\rho_j = \Lambda_j^{-1}(\sigma_i)$ encountered along the way satisfy $\nega(\rho_j) \leq 1$. In this case the input states do not need to satisfy $\nega(\rho) \leq 1$ --- it is only required that $\Tr(\sigma_f \rho)$ can be computed efficiently.

The operators encountered in the middle of the circuit always take the form $\Lambda(\sigma_i)$ (or $\Lambda^{-1}(\sigma_i)$). To bound the cost of an operation one can calculate $\max_i \nega(\Lambda(\sigma_i))$. For Clifford gates and other stabilizer channels we clearly have $\max_i \nega(\Lambda(\sigma_i)) \leq 1$. The protocol can thus efficiently evaluate probability distributions over stabilizer circuits.

Interestingly, stabilizer circuits can be simulated without ever writing down a stabilizer state. In the noise-free case the protocol remains probabilistic, whereas \cite{bennink} becomes deterministic.

In the next section we discuss $\nega(\rho)$, and the conditions under which non-stabilizer quantum states and channels can be efficiently simulated.


\section{Bound states in even-dimensional systems}

The total runtime of the protocol is bounded by $\prod_j \nega(\rho_j)^2$, where the product is over all operators $\rho_j = \Lambda_j(\sigma_i)$ encountered when stepping through the circuit. The cost of each operator is: \begin{equation}
\nega(\rho) = \frac{1}{2^k} \sum_i \abs{ \Tr\pare{  \sigma_i \rho }  }
\end{equation}
States $\rho$ satisfying $\mathcal{D}(\rho)\leq  1$, and observables $E$ satisfying  $\mathcal{D}(E)\leq  1$ are can be efficiently simulated by the algorithm presented in section I. The cost of a channel can be bounded by $\max_i \nega(\Lambda(\sigma_i))$.

Let $\vec q$ be a quasiprobability distribution over stabilizer states $\ket{\phi_i}\bra{\phi_i}$ satisfying $\sum_i q_i = 1$. Define:
\begin{equation}
\mathcal{R}(\rho) = \min_{\vec q} \sum_{i} \abs{x_i} \text{ subject to } \rho = \sum_{i} q_i \ket{\phi_i}\bra{\phi_i} 
\end{equation}

The function $\mathcal{R}$ measures how far $\rho$ is from being a convex combination of stabilizer states.  $\mathcal{R}$ is discussed extensively in \cite{resource}. $\mathcal{D}$ appears in the appendix of \cite{resource}, where they prove $\mathcal{D}(\rho) \leq \mathcal{R}(\rho)$ for all $\rho$. $\mathcal{R}(\rho)$ is the cost of quantum states in the algorithm described by \cite{bennink}.

Together, $\mathcal{R}$ and $\mathcal{D}$ give a classification of multi-qubit quantum states:

\begin{itemize}
    \item \textbf{Stabilizer mixtures} satisfy $\mathcal{R}(\rho) = 1$, $\mathcal{D}(\rho) \leq 1$ and are efficiently simulated by \cite{bennink} and our work.
    \item \textbf{Bound states} satisfy $\mathcal{R}(\rho) > 1$ and $\mathcal{D}(\rho) \leq 1$ and can only be efficiently simulated by this work.
    \item \textbf{Magic states} satisfy $\mathcal{R}(\rho), \mathcal{D}(\rho) > 1$ and cannot be simulated efficiently. 
\end{itemize}

Bound states are well-known to exist in qutrits and other odd-dimensional systems \cite{vcge12} \cite{dh15} \cite{acb12} \cite{hwve14}. We believe our work is the first to point out their existence for multi-qubit systems. Bound states do not exist for the single-qubit system (which might explain why their existence does not appear to be common knowledge).

Figure~\ref{fig:cross-section} shows some cross sections of the 15-dimensional Bloch-space of two-qubit density matrices. Two-qubit bound states are extremely common: Most cross sections feature bound states. We sampled 100,000 random density matrices and found the following distribution:
$$0.9\% \text{ stabilizer},\text{ } 58.3\% \text{ bound},\text{ } 40.8\% \text{ magic}$$

This estimate shows that the number of bound states is much larger than the cross sections would suggest. Estimating the fraction for larger systems is computationally expensive, but it is clear that our algorithm significantly increases the fraction of density matrices that can be efficiently simulated. 
Additionally $\nega(\rho)$ can be as small as $1/2^n$ for the maximally mixed state, meaning that highly mixed states actually \emph{improve} the runtime of the algorithm. This property is not present in \cite{bennink}, or for odd-dimensional systems \cite{pash}.

\section{Case Study}
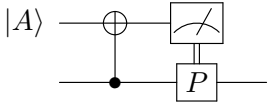
\begin{figure}
    $$P = \begin{bmatrix}1&0\\0&e^{i\pi/2}\end{bmatrix}; \hspace{1cm} T = \begin{bmatrix}1&0\\0&e^{i\pi/4}\end{bmatrix}$$
        $$\ket{A} = \frac{1}{\sqrt{2}}\pare{ \ket{0} + e^{i\pi/4}\ket{1}} = T\ket{+}$$
        \vspace{2mm}
\[
    \Qcircuit @C=1.5em @R=.7em {
        \lstick{\ket{A}} & \targ     & \meter & \\
                      & \ctrl{-1} & \gate{P}\cwx[-1] & \qw
                  }
\]
    \caption{An adaptive Clifford circuit acting on the magic state $\ket{A}$. This `gadget' implements the $T$-gate on the bottom qubit. We denote the channel associated with the Clifford circuit (CNOT and classically-controlled $P$) by $\Lambda$, taking as input two qubits and outputting one qubit. \label{fig:gadget}}

\end{figure}
In this section we describe how to apply our algorithm to a particular circuit (Figure~\ref{fig:gadget}). This circuit is interesting because it illustrates how adaptive Clifford circuits, ancillas and non-stabilizer states together influence the runtime. This circuit is deliberately \emph{not} near-Clifford so the performance will be quite bad.

Direct application of the protocol to this circuit is not efficient, but \emph{not necessary!} A practical implementation of this algorithm \emph{must} include a pre-processing step that uses matrix multiplication to reduce constant-width sub-circuits to a single channel. Pre-composition of channels can only improve runtime:
\begin{equation}\max_i \mathcal{D}(\Lambda_1(\Lambda_2(\sigma_i))) \leq \max_i \mathcal{D}(\Lambda_1(\sigma_i)) \cdot \max_i \mathcal{D}(\Lambda_2(\sigma_i)) \end{equation}

If the circuit above is contracted to a channel we just obtain the $T$-gate. The cost of the $T$ gate is $\max_i \mathcal{D}( T\sigma_i T^\dagger   ) =  \sqrt{2} $, so $n$ repeated applications would increase the runtime by $2^n$. However, $T^n$ reduces to a Clifford gate and at most one $T$ gate.

When forward-propagating through the circuit we must sample a Pauli operator for $\ket{A}$. The we sample $\sigma_i$ with probability $p_i = r_i/\mathcal{D}(\ket{A})$:
$$ \mathcal{D}(\ket{A}) = \frac{1 + \sqrt{2}}{2}; \hspace{5mm} r_I = \frac{1}{2}; \hspace{5mm} r_X = r_Y = \frac{1}{2\sqrt{2}}$$
If we were to apply the algorithm described in \cite{bennink}, the cost would be $\mathcal{R}(\ket{A}) = \sqrt{2}$ which is more expensive than $\mathcal{D}(\ket{A}) = (1+\sqrt{2})/2 \approx 1.207$. However, the adaptive Clifford operation is inexpensive for \cite{bennink}, whereas in our protocol we encounter an additional cost because the operation measures and discards a qubit. If we combine the CNOT and the adaptive $P$-gate into a channel $\Lambda$, we can write:
$$\Lambda(\sigma_{II}) = 2\sigma_I; \hspace{5mm} \Lambda(\sigma_{IZ}) = 2\sigma_Z $$
$$\Lambda(\sigma_{XX}) = \sigma_X + \sigma_Y; \hspace{5mm} \Lambda(\sigma_{XY}) = \sigma_Y - \sigma_X $$
$$\Lambda(\sigma_{YY}) = \sigma_Y - \sigma_X; \hspace{5mm} \Lambda(\sigma_{YX}) = \sigma_X + \sigma_Y $$

Since $\max_i \mathcal{D}( \Lambda(\sigma_i)  ) = 2$, this channel cannot be simulated efficiently and the overall runtime for the circuit when forward-propagating is $\mathcal{D}(\ket{A})^2 \cdot \max_i  \mathcal{D}( \Lambda(\sigma_i)  )^2\approx 5.827$. This cannot be mitigated by backward-propagation since discarding of the ancilla is not due to a local observable but part of the channel $\Lambda$. Calculate $\Lambda^{-1}$:

$$\Lambda^{-1}(\sigma_{I}) = \sigma_{II}; \hspace{5mm} \Lambda^{-1}(\sigma_{Z}) = \sigma_{IZ} $$
$$\Lambda^{-1}(\sigma_{X}) = \bar \sigma_{XX} - \bar\sigma_{XY} - \bar \sigma_{YY} + \bar \sigma_{YX}$$
$$\Lambda^{-1}(\sigma_{Y}) = \bar \sigma_{XX} + \bar\sigma_{XY} + \bar \sigma_{YY} + \bar \sigma_{YX}$$

$\Lambda^{-1}$ is also expensive since $\max_i \mathcal{D}( \Lambda^{-1}(\sigma_i)  ) = 2$. The increased cost of $\Lambda$ and $\Lambda^{-1}$ stems entirely from the discarded qubit. If we append a maximally mixed stabilizer state to $\Lambda$ the cost is removed:  $\max_i \mathcal{D}(\sigma_{I}/2 \otimes  \Lambda(\sigma_i)   ) = 1$. However, the additional state needs to be dealt with later. More generally, since the cost of $n$ maximally mixed states is $\nega(\sigma_I^{\otimes n}/2^n)= 1/2^n$, one might initially think that appending lots of maximally mixed states could make the cost of the protocol arbitrarily small. But discarding these $n$ qubits later costs $2^n$, canceling any cost reduction.

In section II we discussed the cost measure $\mathcal{D}$ acting on states. But if reverse-propagation is employed,  $\mathcal{D}(\rho)$ of input states appears irrelevant. While $\mathcal{D}(\ket{A})$ is not the cost of the circuit in Figure~\ref{fig:gadget}, $\mathcal{D}$ still appears in a different form. In particular if we construct the Choi state of the $T$ gate $\ket{J_T} = (I \otimes T)\ket{\text{Bell}}$ then we see $\mathcal{D}(\ket{J_T}) = \mathcal{D}(\ket{A}) \approx 1.207$.

The cost of an operation, e.g. the $T$ gate, is \emph{not} the cost of its Choi state $\mathcal{D}(\ket{J_T})$. Instead, for an arbitrary channel $\Lambda$ taking $n$ to $m$ qubits, and its Choi state $\ket{J_\Lambda}$:
\begin{equation}
\begin{split}
    \ket{J_\Lambda}\bra{J_\Lambda} &= (\mathcal{I} \otimes \Lambda)\Big[\ket{\text{Bell}}\bra{\text{Bell}}^{\otimes n}\Big]  = \frac{1}{2^n} \sum_{i} \sigma_{i} \otimes \Lambda(\sigma_{i})  \\
    \mathcal{D}(\ket{J_\Lambda}) &= \frac{1}{2^n \cdot 2^{n+m}} \sum_{i} \sum_j \abs{ \Tr((\sigma_{i} \otimes \Lambda(\sigma_{i}))  \sigma_j )} \\
    = \frac{1}{2^{n+m}} &\sum_{i} \sum_{k} \abs{ \Tr( \Lambda(\sigma_{i}) \sigma_k) } = \frac{1}{2^n} \sum_{i} \mathcal{D}(\Lambda(\sigma_i)) 
\end{split}
\end{equation}

So in fact $\mathcal{D}(\ket{J_\Lambda})$ is the \emph{average} of $\mathcal{D}(\Lambda(\sigma_i))$, but the runtime depends on the maximum of $\mathcal{D}(\Lambda(\sigma_i))$. We see that this distinction also manifests in the circuit in Figure~\ref{fig:gadget}: although $\ket{A}$ is a magic state for the $T$-gate, $\mathcal{D}(\ket{A})$ measures the average cost of the gate, not the maximum cost.

\section{Conclusions}

The stabilizer formalism simulates quantum states by embedding symmetry into Hilbert space. Quasiprobability techniques describe the states and operations that belong to that symmetry as the convex polytope of a constellation of hermitian operators. Stabilizer states form a complicated constellation with very many vertices. However, their underlying symmetry group is described by a much simpler constellation: the Pauli matrices. The stabilizer states are just where the Pauli matrix polytope described by $\mathcal{D}$ happens to intersect with the non-linear $\rho^2 = \rho$ purity condition.

This intuitive picture was previously known for odd-dimensional systems \cite{gross}. The phase point operators of the discrete Wigner function form a simplex in Bloch-space, and some of its (hyper-)faces have a stabilizer state at the center. There, the Clifford group is the subgroup of the rotational symmetry group of the simplex expressible with unitaries. For even-dimensional systems the Pauli matrices form a hyper-octahedron. The difference in geometry raises the maximum trace of the operators, resulting in a surmountable complication with discarding qubits that makes reverse-propagation sometimes necessary.

Additive-error probability estimation may be insufficient in some situations. Exponentially small probabilities can not be estimated, only upper-bounded. Sampling from an indistinguishably close distribution is also not possible. Recent work \cite{sampling} explores the connection between probability estimation of various precisions and sampling. \cite{gosset} achieves L1-close sampling for noiseless circuits in the Clifford+$T$ gate set with remarkable efficiency. The question of whether multiplicative-error probability estimation is possible for noisy near-Clifford quantum circuits is extremely important. We caution that if the efficiently simulable circuits for such a protocol are universal for classical computation, then it could be used to efficiently solve NP-complete problems\footnote{Multiplicative-error probability estimation can distinguish probabilities from zero. Consider a randomized algorithm that samples a random proof for an NP question, and accepts only if the proof checks out. The NP problem can be solved by determining if the acceptance probability is nonzero.}.

The Clifford group encompasses many gates commonly used in quantum circuits. However, it is possible that for some noisy circuits a different symmetry group is more appropriate. Furthermore, \cite{zhusic} highlights a trade-off between the size of the symmetry group and the maximum negativity of a quasiprobability representation. An interesting avenue could be to develop a scheme to automatically identify the optimal quasiprobability representation to simulate an input circuit.

Large noisy quantum circuits will be an important subject during the NISQ era \cite{nisq}. Since classical examination of quantum devices is an indispensable tool for development of scalable quantum computation, we expect that symmetry in Hilbert space will prove to be a powerful ally.

\section{Acknowledgments}

I would like to thank Dr. Scott Aaronson (UT Austin) for supporting my work and giving me helpful advice and encouragement. I am also grateful for helpful conversations with Dr. James Troupe (Applied Research Labs, Austin), who also pointed me to some useful literature.

\end{document}